\documentclass[reprint,twocolumn,pre,showpacs,amsmath,amssymb,aps]{revtex4-1}

\usepackage{natbib}	
\usepackage{graphicx,color,rotating}
\usepackage[latin1]{inputenc}
\usepackage{textcomp}
\usepackage{dcolumn}
\usepackage{bm}     
\usepackage{upgreek}
\usepackage{subdepth}  			
\usepackage[latin1]{inputenc}

\usepackage{array}
\newcolumntype{L}[1]{>{\raggedright\let\newline\\\arraybackslash\hspace{0pt}}m{#1}}
\newcolumntype{C}[1]{>{\centering\let\newline\\\arraybackslash\hspace{0pt}}m{#1}}
\newcolumntype{R}[1]{>{\raggedleft\let\newline\\\arraybackslash\hspace{0pt}}m{#1}}

\usepackage{hyperref}
\usepackage{xcolor}							
\hypersetup{							    
    colorlinks,
    linkcolor={blue!80!black},
    citecolor={blue!80!black},
    urlcolor={blue!80!black}
}

\newcommand{\mr}[1]{\ensuremath{\mathrm{#1}}}
\renewcommand{\vec}[1]{\bm{#1}}
\newcommand{\ee}{\mathrm{e}}
\newcommand{\ii}{\mathrm{i}}
\newcommand{\dm}{\mathrm{d}}

\newcommand{\avr}[1]{\big\langle #1 \big\rangle}

\newcommand{\iot}{{\ii\omega t}}

\newcommand{\ve}{\varepsilon}

\newcommand{\pp}{\partial}
\newcommand{\ppsqr}{\partial^{\,2_{}}}

\newcommand{\nablabf}{\boldsymbol{\nabla}}

\newcommand{\Lapl}{\nabla^2}

\newcommand{\divop}{\nablabf\cdot}

\newcommand{\scap}{\!\cdot\!}

\newcommand{\fracsmall}[2]{\mbox{$\frac{#1}{#2}$}}

\newcommand{\ddd}{\vec{d}}

\newcommand{\FFFrad}{\vec{F}^\mathrm{rad}}

\newcommand{\fffac}{\vec{f}_\mathrm{ac}}

\newcommand{\gvec}{\vec{g}}

\newcommand{\nnn}{\vec{n}}

\newcommand{\rrr}{\vec{r}}
\newcommand{\rrrO}{\vec{r}_0}

\newcommand{\uuu}{\vec{u}}

\newcommand{\vvv}{\vec{v}}

\newcommand{\zerovec}{\boldsymbol{0}}

\newcommand{\cO}{c_0}

\newcommand{\Dth}{D_\mathrm{th}}

\newcommand{\kth}{k_\mathrm{th}}

\newcommand{\kapT}{\kappa_T}

\newcommand{\pii}{p_{11}}                  

\newcommand{\alfP}{{\alpha_p}}
\newcommand{\alfpTi}{\tilde{\alpha}_p}

\newcommand{\delt}{\delta_\mathrm{t}}

\newcommand{\dels}{\delta_\mathrm{s}}

\newcommand{\etaO}{\eta_0}

\newcommand{\etaOTi}{\tilde{\eta}_0}

\newcommand{\sigmabfF}{\sigmabf_\mathrm{f}}
\newcommand{\sigmabfS}{\sigmabf_\mathrm{s}}
\newcommand{\sigmabfIF}{\sigmabf_\mathrm{1f}}
\newcommand{\sigmabfIS}{\sigmabf_\mathrm{1s}}

\newcommand{\cOsqr}{c^{\,2_{}}_0}

\newcommand{\cP}{c_\mathrm{p}}

\newcommand{\fOfl}{f_0^\mathrm{fl}}

\newcommand{\fIfl}{f_1^\mathrm{fl}}

\newcommand{\kapO}{\kappa_0}

\newcommand{\pO}{p_0}

\newcommand{\pI}{p_1}

\newcommand{\pII}{p_2}

\newcommand{\TO}{T_0}
\newcommand{\TI}{T_1}
\newcommand{\TII}{T_2}

\newcommand{\vvvI}{\vvv_1}

\newcommand{\vvvII}{\vvv_2}

\newcommand{\rhoO}{\rho_0}

\newcommand{\rhoI}{\rho_1}
\newcommand{\rhoII}{\rho_2}

\newcommand{\rhoOTi}{\tilde{\rho}_0}

\newcommand{\SImum}{\textrm{\textmu{}m}}

\newcommand{\SIms}{\textrm{ms}}
\newcommand{\SImus}{\textrm{\textmu{}s}}

\newcommand{\beq}[1]{\begin{equation} \eqlab{#1}}
\newcommand{\eeq}{\end{equation}}
\newcommand{\bsub}{\begin{subequations}}
\newcommand{\esub}{\end{subequations}}
\def\bal#1\eal{\begin{align}#1\end{align}}
\def\bsubal#1\esubal{\bsub \begin{align}#1\end{align} \esub}

\newcommand{\eqlab}[1]{\label{eq:#1}}
\renewcommand{\eqref}[1]{Eq.~(\ref{eq:#1})}
\newcommand{\eqnoref}[1]{(\ref{eq:#1})}

\newcommand{\xs}{x_\mathrm{s}}
\newcommand{\xt}{x_\mathrm{t}}

\newcommand{\xsp}{x_\mathrm{s}^{\prime}}
\newcommand{\xtp}{x_\mathrm{t}^{\prime}}

\newcommand{\Pibf}{\bm{\Pi}}
\newcommand{\Piac}{\bm{\Pi}_\mathrm{ac}}
\newcommand{\sigmabf}{\bm{\sigma}}
\newcommand{\cL}{c_\mathrm{L}}
\newcommand{\cT}{c_\mathrm{T}}
\newcommand{\cTsqr}{c^2_\mathrm{T}}
\newcommand{\cLsqr}{c^2_\mathrm{L}}

\newcommand{\uuuI}{\vec{u}_1}

\newcommand{\sigmabfII}{\bm{\sigma}^{{}}_2}

\begin{document}

\title{Theoretical aspects of microscale acoustofluidics}

\author{Henrik Bruus}
\affiliation{Department of Physics, Technical University of Denmark, DTU Physics Building 309, DK-2800 Kongens Lyngby, Denmark}

\date{27 February 2017. E-mail: bruus@fysik.dtu.dk}

\begin{abstract}
\vspace*{1mm} \noindent
\begin{tabular}{l@{\hskip 0.0mm}l}
\framebox{\includegraphics[width=21mm]{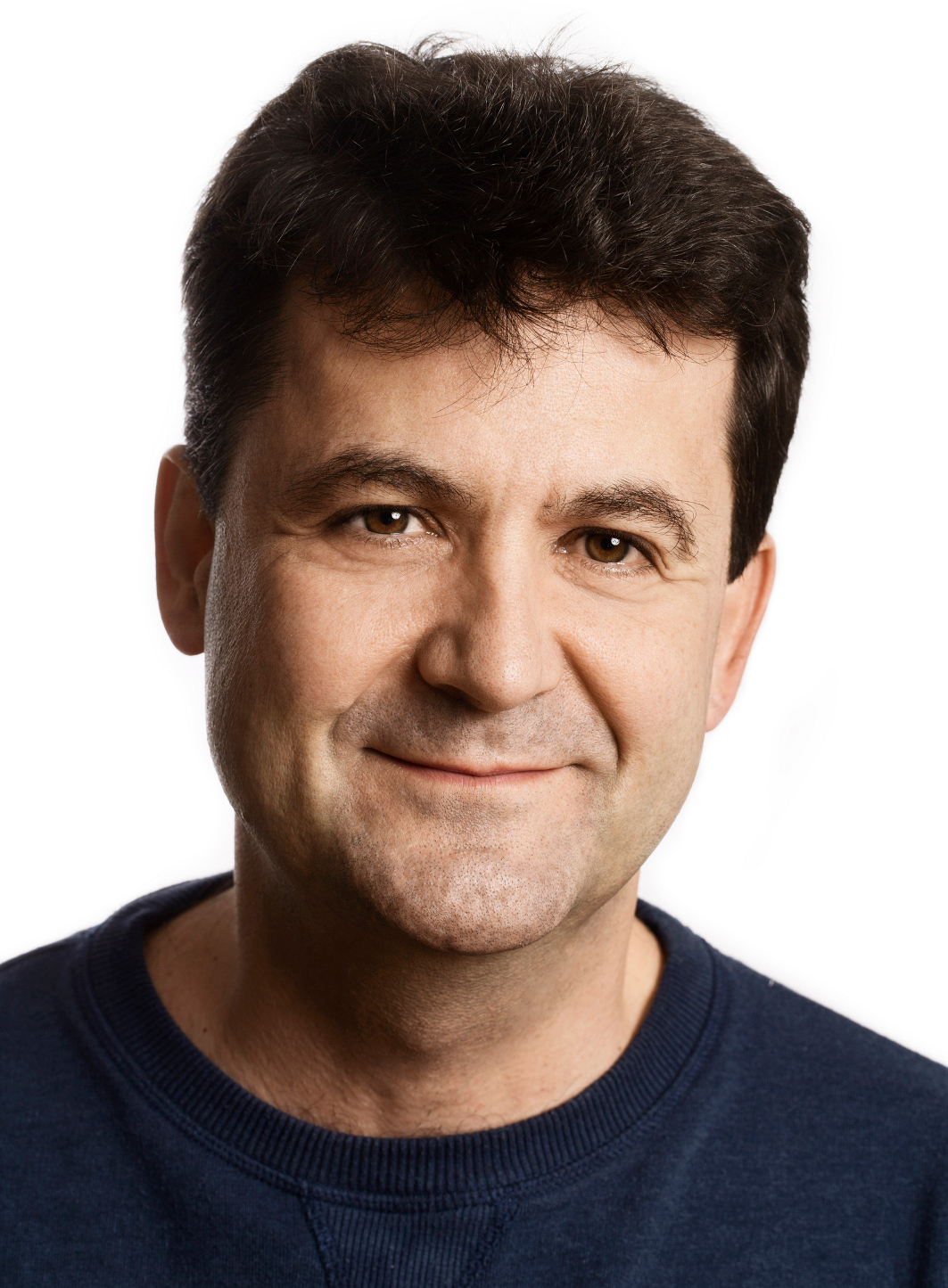}} \rule[0.0mm]{0mm}{0mm}
&
\begin{minipage}[b][30mm][l]{115mm}
\textbf{Short biography:}  Henrik Bruus received his PhD degree in physics from the Niels Bohr Institute, University of Copenhagen in 1990, and then worked as postdoc at Nordic Institute of Theoretical Physics 1990-92, Yale University 1992-94 and CNRS Grenoble 1994-96. He returned to the Niels Bohr Institute as associate professor 1997-2001, before moving to the Technical University of Denmark in 2001. There, he became full professor of lab-chip systems in 2005 and of theoretical physics in 2012. His current research interests comprise micro/nanofluidics, acoustofluidics, electrokinetics, the physics of on-chip cell manipulation, the motion of sugar in
living plants, and topology-optimized microflows. He has (co)au- \end{minipage}
\end{tabular}\\[1.5mm]
thored  134 journal papers on condensed matter physics and microfluidics, 180 conference papers, and 2 monographs, the latest being "Theoretical Microfluidics", Oxford University Press (2008).\\[4mm]
\textbf{Abstract}.
In this contribution, I summarize some of the recent results within theory and simulation of microscale acoustofluidic systems that I have obtained in collaboration with my students and international colleagues. The main emphasis is on three dynamical effects induced by external ultrasound fields acting on aqueous solutions and particle suspensions: The acoustic radiation force acting on suspended micro- and nanoparticles, the acoustic streaming appearing in the fluid, and the newly discovered acoustic body force acting on inhomogeneous solutions.
\end{abstract}




\maketitle

\section{Introduction}

Microscale acoustofluidic devices are used increasingly in biology, environmental and forensic sciences, and clinical diagnostics as reviewed in the Lab Chip Tutorial Series~\cite{Bruus2011c}, reprinted  by the Royal Society of Chemistry in the book \textit{Microscale Acoustofluidics}~\cite{Laurell2014}.

Following the standard method described in textbooks such as Lighthill \cite{Lighthill2002}, Pierce \cite{Pierce1991}, and Landau \& Lifshitz \cite{Landau1993}, we formulate the linear acoustics in fluids and elastic solids, as well as the non-linear acoustic dynamics in fluids using first-and second-order perturbation theory in the notation established in the textbook by Bruus~\cite{Bruus2008}. The basic fields in both fluids and solids are the density $\rho$, the stress tensor $\sigmabf$, the internal energy density $\ve$, and the temperature $T$. These fields are supplemented in the fluids by the pressure $p$ and the velocity $\vvv$, and in the solids by the displacement $\uuu$.

\section{Governing equations}

For fluids, advection causes non-linearities, and the governing equations are formulated in terms of the fluxes of the mass density $\rho$, the momentum density $\rho\vvv$, and the energy density $\rho\big(\ve + \fracsmall{1}{2}v^2\big)$ under the simplifying assumption of no net body forces and no heat sources,
 \bsub
 \eqlab{governing_equations}
 \bal
 \eqlab{contEq}
 \pp_t \rho &= \nablabf\scap\big[-\!\rho\vvv\big] , \\
 \eqlab{momentumEq}
 \pp_t (\rho\vvv) &= \nablabf\scap\big[\sigmabfF  - \rho\vvv\vvv\big] , \\
 \eqlab{energyEq}
 \pp_t \big(\rho\ve\!+\!\fracsmall{1}{2}\rho v^2\big) &=
 \nablabf\scap\big[\kth\nablabf T +\! \vvv\scap\sigmabfF -\! \rho(\ve+\fracsmall{1}{2}v^2)\vvv\big],
 \eal
 \esub
where the stress tensor $\sigmabfF$ for fluids is
 \beq{sigmaFdef}
 \sigmabfF  = -\Big[p \!+\! \big(\fracsmall{2}{3} \eta_0 - \eta_\mr{B}\big)\nablabf\scap\vvv \Big]\vec{1}
 + \etaO\!\Big[\nablabf\vvv +(\nablabf\vvv)^\mathrm{T}\:\Big].
 \eeq
Here, the material parameters are the shear viscosity $\etaO$, the bulk viscosity $\eta_\mr{B}$, and the thermal conductivity $\kth$, while $\vec{1}$ is the unit tensor and 'T' means tensor transpose.

For elastic solids with an isobaric thermal expansion coefficient $\alfP$, an isothermal compressibility $\kapT$, and a specific heat ratio $\gamma$, the governing equations are linear due to the lack of advection,
 \bsubal
 \eqlab{u_solid_equ}
 \rho\ppsqr_t \uuu &= \nablabf\scap\sigmabfS -\frac{\alfP}{\kapT}\:\nablabf T,\\
 \eqlab{T_solid_equ}
 \pp_t T &= \gamma\Dth\Lapl T -\frac{\gamma-1}{\alfP}\pp_t\nablabf\scap\uuu,
 \esubal
where the stress tensor $\sigmabfS$ for solids is given in terms of the longitudinal and transverse sound speeds $\cL$ and $\cT$, respectively,
 \beq{sigmaSdef}
 \sigmabfS =  \rho\Big[\cLsqr-2\cTsqr\Big](\nablabf\scap\uuu)\vec{1}
 + \rho\cTsqr\!\Big[\nablabf\uuu + (\nablabf\uuu)^{\mathrm{T}}\:\Big].
 \eeq

\section{Acoustic perturbation theory}

We consider a system initially at rest, and which has the density $\rhoO$ and velocity $\vvv = \zerovec$. We assume that the system is perturbed by imposing an harmonically oscillating displacement with a small position-dependent amplitude $\ddd$ on parts of the boundary. As a result, harmonic variations $\rhoI$ in the density and $\vvvI$ in the velocity are induced. For a medium with the speed of sound $\cO$, the magnitude of the perturbation can then be characterized by the dimensionless acoustic Mach number Ma,
 \beq{MaDef}
 \text{Ma} = \frac{v_1}{\cO} = \frac{\rhoI}{\rhoO} \ll 1.
 \eeq
Typically in microscale acoustofluidics $\text{Ma} \approx 10^{-4}$, so a perturbation expansion of any field $f$ in order of Ma makes sense, $f = f_0 + f_1 + f_2 + \ldots$, where $f_n \propto \text{Ma}^n$. We expand the non-linear fluid equations to second order, while the linear solid equations terminate at first order,
 \beq{AcoustExpand2}
 \begin{array}{rll}
 T &= \TO + \TI + \TII, & \quad \uuu = \zerovec + \uuuI,\\[1mm]
 \rho &= \rhoO + \rhoI + \rhoII,  & \quad \vvv = \zerovec + \vvvI+ \vvvII,\\[1.5mm]
 p &= \pO + \pI + \pII. & \end{array}
 \eeq

\subsection{The first-order equations}
Using the complex-valued representation $\ee^{-\iot}$ of harmonic time dependence with angular frequency $\omega$, the actuation of the system is modeled by prescribing the motion of its exterior boundaries at positions $\rrrO$,
 \bsub
 \begin{alignat}{2}
 \eqlab{BCs1}
 \uuuI &= &\ddd(\rrrO)\:\ee^{-\iot},&\; \text{ on solid boundaries},\\
 \vvvI &=\; &-\ii\omega\ddd(\rrrO)\:\ee^{-\iot},&\; \text{ on  fluid boundaries}.
 \end{alignat}
 \esub
All first-order fields thus have the form $g_1 = g_1(\rrr)\:\ee^{-\iot}$, which changes the time derivative into a simple complex-valued factor, $\pp_t \rightarrow -\ii \omega$, and consequently, all first-order  terms contain the harmonic phase-factor $\ee^{-\iot}$. This factor is divided out leaving the time-independent, complex-valued amplitude fields $g_1(\rrr)$.

For fluids in the isentropic case, we have the simplifying relations $\pI = \cOsqr\rhoI$ and $\TI = \frac{\alfP\TO}{\cP\rhoO}\:\pI$, and the governing equations for the first-order fields become
 \bsub
 \eqlab{FirstOrderFluid}
 \bal
 \eqlab{CEequ1}
 -\ii\omega\pI &= -\rhoO\cOsqr\:\nablabf\cdot\vvvI,\\
 \eqlab{NSequ1}
 -\ii\omega\rhoO\vvvI &= \nablabf\cdot\sigmabfIF,
 \eal
 \esub
where the first-order stress tensor $\sigmabfIF$ is \eqref{sigmaFdef} with all fields substituted by first-order fields. Similarly for the solids, we obtain the simplified first-order equation of motion,
 \beq{FirstOrderSolid}
 -\omega^2\rhoO\uuuI = \nablabf\cdot\sigmabfIS.
 \eeq

At internal fluid-solid interfaces with the surface normal vector $\nnn$, we demand continuity of the velocity and stress fields,
 \beq{FirstOrderBCfs}
 \left. \begin{array}{rl}
 \vvvI =& -\ii\omega\uuuI\\[1mm]
 \sigmabfIF\scap\nnn =&  \sigmabfIS\scap\nnn
 \end{array}\!\! \right\}\!,
 \text{internal fluid-solid interfaces}.
 \eeq

\subsection{The time-averaged second-order equations}

The basic acoustofluidic dynamics takes place in fluids on a ms-time scale, much slower than the ultrasonic $\SImus$-time scale. It is therefore described by the time-averaged response over one oscillation period, denoted by angled brackets $\langle \ldots \rangle$ in the following. The terms of lowest order contributing to this time average are second-order terms. Moreover, since the time-average of the time derivative $\avr{\pp_t g}$ of any field $g$ in time-periodic systems vanish, $\langle \pp_t g \rangle = 0$, the second-order governing equations of fluids become
 \bsubal
 \eqlab{CEequ2}
 \nablabf\cdot\Big[\rhoO\avr{\vvvII} + \avr{\rhoI\vvvI}\Big] &= 0,\\
 \eqlab{NSequ2}
 \nablabf\cdot\Big[\avr{\sigmabfII} - \rhoO\avr{\vvvI\vvvI}\Big] &= \zerovec.
 \esubal
In terms of the momentum flux density tensor
 \beq{PiDef}
 \Pibf = \rho\vvv\vvv - \sigmabf,
 \eeq
the second-order equation \eqnoref{NSequ2} becomes
 \beq{Pi2equ}
 \nablabf\cdot\avr{\Pibf_2} = \zerovec.
 \eeq

On the boundary, the velocity is oscillating harmonically, so there $\avr{\vvv} = \zerovec$. On the other hand, a boundary point with the time-averaged position $\rrrO$ has the time-dependent position $\rrr = \rrrO + \uuuI$, so by a second-order Taylor expansion $\avr{\vvv(\rrr)} = \avr{\vvv(\rrrO) + \uuuI(\rrrO)\scap\nablabf\vvv(\rrrO)}$. Thus on the boundary $\zerovec = \avr{\vvvII}+\avr{\uuuI\cdot\vvvI}$, and using $\uuuI = \frac{\ii}{\omega}\vvvI$ from \eqref{FirstOrderBCfs}, the boundary condition for $\avr{\vvvII}$ becomes
 \beq{v2BC}
 \avr{\vvvII} = -\frac{1}{\omega}\avr{(\ii\vvvI)\cdot\nablabf\vvvI}, \text{ on fluid boundaries}.
 \eeq

\section{The radiation force}

In a series of papers, we have studied the acoustic radiation force $\FFFrad$ on a spherical particle of radius $a$ suspended in an inviscid \cite{Bruus2012a}, viscous \cite{Settnes2012}, and thermoviscous fluid \cite{Karlsen2015} with kinematic viscosity $\nu_0$ and thermal diffusivity $\Dth$. When exposed to an incident acoustic wave of angular frequency $\omega$ and wavenumber $k$, a viscous and thermal boundary layer of width $\dels=\sqrt{2\nu_0/\omega}$ and $\delt=\sqrt{2\Dth/\omega}$, respectively, develop at the particle surface. In the long-wave limit $a,\dels,\delt \ll 1/k$, we can express $\FFFrad$ as the time-averaged momentum flux through any static surface $\pp\Omega_0$ encompassing the particle,
 \beq{Frad}
 \FFFrad = -\left\langle \oint_{\pp\Omega_0}\!\!\Pibf_2\cdot\nnn\:\dm a\right\rangle
 = -\int_{\Omega_0}\!\!\nablabf\cdot\avr{\Pibf_2}\:\dm V.
 \eeq
In spite of \eqref{Pi2equ}, the expression for $\FFFrad$ is not zero because of delta-function singularities which appear in $\nablabf\cdot\avr{\Pibf_2}$ due to the scattering of the incident acoustic wave on the particle. When properly evaluating these singularities we find \cite{Settnes2012},
 \bal
 \eqlab{Frad_Settnes}
 \FFFrad \!= - \frac43\pi a^3\! \Big[\kapO\avr{(f_0 p_1^\mathrm{in})\! \nablabf p_1^\mathrm{in}} \big]
 - \frac32\rho_0\avr{(f_1 \vec{v}_1^\mathrm{in}) \!\cdot\! \nablabf \vec{v}_1^\mathrm{in}}\Big],
 \eal
where $p_1^\mathrm{in}$ and $\vec{v}_1^\mathrm{in}$ are the first-order pressure and velocity fields of the incident acoustic wave evaluated at the particle position,  $\kapO$ is the isentropic compressibility  of the fluid, while $f_0$ and $f_1$ are the so-called mono- and dipole scattering coefficients, respectively. The latter are involved functions of the material parameters as shown in Ref.~\cite{Karlsen2015}. For a fluid particle (a droplet), $f_0$ becomes
 \bsub
 \eqlab{f0_fluid-fluid}
 \bal
 &\fOfl = 1 - \tilde{\kappa}_s + 3 (\gamma - 1 )
 \left(1 - \dfrac{ \alfpTi }{\rhoOTi \: \tilde{c}_p} \right)^2 H(\xt,\xtp),
 \\
 \eqlab{monopole_H}
 &H(\xt,\xtp) = \dfrac{1}{\xt^2} \left[ \dfrac{1}{1-\ii \xt} - \dfrac{1}{\tilde{k}_\mathrm{th}} \dfrac{\tan \xt^{\prime}}{\tan \xt^{\prime} - \xt^{\prime}} \right]^{-1},
 \eal
 \esub
where the tilde represent the droplet-to-medium ratio of the various material parameters, and where $H(\xt,\xtp)$ is a function of the particle radius $a$ through the non-dimensionalized thermal wavenumber $\xt\approx(1+\ii)a/\delt$ for the fluid medium, and likewise for $\xt^\prime$ of the droplet.

Similarly, $f_1$ becomes
 \bsub
 \eqlab{f1_fluid-fluid}
 \bal
 \eqlab{f1_fl_final}
 \fIfl &= \dfrac{2\left(\tilde{\rho_0}-1\right)\left(1+F(\xs,\xsp) -G(\xs) \right)}
 {\left(2\tilde{\rho_0}+1\right)\big[1+F(\xs,\xsp)\big]-3 G(\xs)} ,
 \\[2mm]
 \eqlab{G_def}
 G(\xs) &= \dfrac{3}{\xs} \left(\dfrac{1}{\xs} - \ii \right),
 \\[2mm]
 \eqlab{F_def}
 F(\xs,\xsp) & = \dfrac{1 - \ii \xs}{2 (1- \etaOTi ) + \dfrac{\etaOTi \xs^{\prime 2} (\tan \xs^{\prime} - \xs^{\prime}
 )}{(3-\xs^{\prime 2})\tan \xs^{\prime} - 3 \xs^{\prime}}},
  \eal
  \esub
where $\xs=(1+\ii)a/\dels$ is the non-dimensionalized shear-viscous wavenumber for the fluid medium, and likewise for $\xs^\prime$ of the droplet.

A full experimental verification of our theoretical expression for the acoustic radiation force has not yet been carried out. However, we have successfully obtained partial experimental verifications~\cite{Barnkob2010, Augustsson2011, Barnkob2012a}.\\[-6mm]

\section{Acoustic streaming in homogeneous fluids}

\vspace*{-2mm}
In a series papers, we have studied theoretically and numerically the acoustic streaming in hard-walled acoustofluidic devices \cite{Muller2012, Muller2013, Muller2014, Antfolk2014, Muller2015}. All types of time-averaged acoustic flows are driven by an acoustic body force $\fffac$, which is given in terms of the non-zero divergence of the acoustic part $\avr{\Piac}$ of the momentum-flux-density tensor $\avr{\Pibf_2}$. We define $\fffac$ as~\cite{Karlsen2016}
 \bsub
 \eqlab{facGeneral}
 \beq{facDef}
 \fffac = -\divop\avr{\Piac},
 \eeq
where $\avr{\Piac}$ is the part of $\avr{\Pibf_2}$ that contains all the time-averaged products of first-order acoustic fields,
 \bal
 \eqlab{Piac}
 \avr{\Piac} &= \avr{ \pii } \mathbf{I} + \avr{\rhoO\vvvI\vvvI},
 \quad\quad \mathrm{with} \\
 \avr{ \pii } &= \avr{ \frac12 \kapO |\pI|^2 - \frac12 \rhoO |\vvvI|^2 } .
 \eal
 \esub
In this expression, $\avr{\pii}$ is a local time-averaged acoustic pressure. Using the body force $\fffac$, we rewrite \eqref{NSequ2} as an equation for a body-force-driven Stokes flow,\\[-6mm]
 \beq{StokesEqu2}
 \nablabf\cdot\avr{\sigmabfII} = \fffac.\\[-2mm]
 \eeq
For a planar rigid wall and a homogeneous fluid, we recover the well-known Rayleigh streaming, with pairs of vortices and anti-vortices each with a spatial extension of $\frac14\lambda \times \frac14\lambda$ \cite{Muller2012}. In the same work we added tracer particles and studied the transition in particle dynamics from being dominated by the radiation force to be dominated by the Stokes drag force from the acoustic streaming. We determined the critical radius $a_0$ for the cross-over between these two regimes to be around $1~\SImum$ for polystyrene particles in water at 2~MHz \cite{Muller2012, Barnkob2012}.

In subsequent studies of rectangular microchannels, we compared the theoretical predictions of our streaming model with experimental observations for the standard case of a horizontal standing half-wave acoustic resonance. In Ref.~\cite{Barnkob2012} we verified experimentally the model for the cross-over from radiation-force to streaming-drag dominated particle dynamics, and in Ref.~\cite{Muller2013} we found good agreement between experimental, analytical, and numerical determinations of streaming-dominated particle paths in all three spatial dimensions. Finally in Ref.~\cite{Antfolk2014}, we found good agreement between theory and experiments for the handling of sub-$\SImum$ particles in the special case of single-vortex streaming obtained by phase-specific oscillation of both pairs of opposite walls in a rectangular channel.\\[-6mm]

\section{Acoustic body-force handling of inhomogeneous solutions}

\vspace*{-2mm}
Recently, it was discovered experimentally that inhomogeneities in density $\rhoO$, compressibility $\kapO$, and viscosity $\etaO$ of a fluid, introduced by a solute concentration $s$, can be relocated and stabilized into field-dependent configurations by use of acoustic fields~\cite{Deshmukh2014, Augustsson2016}. Based on the separation of the fast $\SImus$-time-scale $t$ of acoustics and the slow $\SIms$-time-scale $\tau$ of hydrodynamics, we subsequently developed the following theory~\cite{Karlsen2016} for the acoustic body force responsible for the observed phenomena.

The density $\rho$ is written as the sum of the slowly evolving hydrodynamic density $\rhoO$ and the quickly oscillating density perturbation $\rhoI$,
 \beq{PertExpansion}
 \rho = \rhoO(\rrr,\tau) + \rhoI(\rrr,\tau)\:\ee^{-\iot},
 \eeq
where $\rhoI$ is associated with the acoustic pressure and velocity fields $\pI$ and $\vvvI$. Now, the two equations for $\pI$ and $\rhoI$ cannot be combined into one as in \eqref{FirstOrderFluid}, so
inhomogeneous acoustics are given by three equations,
 \bsub
 \eqlab{FirstOrderEqs}
 \bal
 - \ii \omega \kapO \pI &= - \divop\vvvI, \\
 - \ii \omega \rhoO \kapO \pI &= - \ii \omega \rhoI + \vvvI \cdot \nablabf \rhoO,\\
 - \ii \omega \rhoO \vvvI &= \divop\sigmabfIF.
 \eal
 \esub

The fluid inhomogeneity is caused by a solute concentration field $s=s(\rrr,\tau)$, which is being transported on the slow timescale $\tau$. The density $\rhoO\big(s(\rrr,\tau)\big)$, compressibility $\kapO\big(s(\rrr,\tau)\big)$, and viscosity $\etaO\big(s(\rrr,\tau)\big)$ are all functions of the solute concentration, and consequently the hydrodynamics is governed by the Stokes equation with body forces (acoustics and gravity) for the velocity field $\vvv$, the continuity equation for $\rhoO$, and the advection-diffusion equation of the solute $s$,
 \bsub
 \eqlab{DynamicsSlow}
 \bal
 \eqlab{NSSlow}
 \pp_\tau (\rhoO \vvv) &= \divop \sigmabf + \fffac + \rhoO \gvec , \\
 \eqlab{ContSlow}
 \pp_\tau \rhoO &= - \divop \big( \rhoO \vvv \big) , \\
 \eqlab{DiffusionSlow}
 \pp_\tau s &= - \divop \big[ - D \nablabf s + \vvv s \big] .
 \eal
 \esub

Usual streaming in the homogeneous viscous case \cite{Muller2012, Muller2013} is obtained using $s=0$ and $\fffac$ of the form
 \beq{facHom}
 \fffac^\mathrm{hom} = - \rhoO\divop\avr{\vvvI\vvvI}.
 \eeq

The recently discovered slow-time-scale relocation of solute inhomogeneities into stable field-dependent configurations is obtained in the inhomogeneous inviscid case, for which $s\neq0$ and and $\fffac$ reduces to \cite{Karlsen2016, Karlsen2017}
 \beq{facInv}
 \fffac^\mathrm{invisc} = - \frac14 |\pI|^2 \nablabf\kapO(s) - \frac14 |\vvvI|^2 \nablabf\rhoO(s).
 \eeq

In the general inhomogeneous viscous case, which we are currently studying, the streaming flow co-develop with the solute relocation using $\fffac$ from \eqref{facGeneral},
 \beq{facGeneral2}
 \fffac = -\nablabf\avr{p_{11}(s)} - \nablabf\cdot\avr{\rhoO(s)\vvvI\vvvI}.
 \eeq

\section{Discussion}
Central aspects and selected examples of these recent theoretical developments in the field of microscale acoustofluidics will be discussed at the Nobel Symposium on Microfluidics, Sånga-S{\"a}by, Sweden, 5 -- 8 June 2017.

%
%


%

\end{document}